# Simultaneously plasmon lasing and spasing behavior in a silver grating-film geometry

Lina Shi,[1] Hailiang Li,[1] Feng Jin,[2] Jiebin Niu,[1] Yilei Hua,[1] and Changqing Xie[1, a)]

[1]*Key Laboratory of Microelectronics Devices & Integrated Technology, Institute of Microelectronics, Chinese Academy of Sciences, Beijing 100029, China*

[2]*Laboratory of Organic NanoPhotonics and Key Laboratory of Functional Crystals and Laser Technology, Technical Institute of Physics and Chemistry, Chinese Academy of Sciences, Beijing, 100190, China*

By using a self-consistent Maxwell-Bloch method, we demonstrate the simultaneously lasing and spasing behavior in a simple metal grating-film nanostructure, which can be attributed to spatial hole burning and the gain competition of different modes at the band edge and in the plasmonic band gap. We show three modes: one spaser mode in gap with quality factor as high as 248.54, one plasmon lasing mode at band edge which emit vertically from the grating surface, and the other plasmon lasing mode at band edge which is suppressed by the spaser mode. This method may find significant applications in coherent light and surface plasmon sources with low threshold, surface enhanced Raman scattering, solid-state lighting emission, etc.

---

a) Author to whom correspondence should be addressed.

Electronic mail: xiechangqing@ime.ac.cn

Spaser, a quantum amplifier of surface plasmons based on stimulated emission of radiation, has attracted tremendous attention since it was first proposed by Bergman and Stockman.[1-2] The system consists of a metal nanoplasmonic component and a nanoscale gain medium, such as dye molecules or quantum dots, where the population inversion is created optically or electrically. The basic operating principle of the spaser is similar to that of laser, with the role of coherent photons replaced by surface plasmons (SPs), and the optical resonant cavity replaced by metal nanoparticles with deep subwavelength confinement. The idea of a spaser has developed rapidly, and spasers have been experimentally demonstrated in various geometries with one-, two-, and three-dimensional confinement.[3-10] Many interesting optical characteristics such as intensity-dependent frequency shift, line width and Rabi oscillations have been reported.[11-17] And compensating joule loss in metamaterials is considered as the most promising application of spasers.[18-20]

In a further development of the spaser concept, the lasing spaser was suggested to power up a nanoscale source of coherent radiation.[21] Different from the dark mode (does not emit photons) of the spaser, lasing spaser utilize bright modes, such as the trapped modes[22] or toroidal dipolar modes[23] supported by an array of plasmonic nanoparticles, and emit electromagnetic radiations to the far field. However, the reported spasers lack the control of the emission directionality, and the spasing and lasing behavior in a single platform hasn't been reported yet. Furthermore, the reported spasers or lasing spasers were mostly implemented by nanoparticle or nano-gap geometry where SP was confined in specific three spatial dimensions, and such configurations need extra structures in z-direction. These weaknesses limit the ability to fabricate high quality and reliable spasers or lasing spasers, and to integrate with other on-chip passive optical or plasmonic components.

In this letter, we demonstrate the simultaneously spasing and lasing behaviors in a silver grating-film geometry. The grating-film geometry, as shown in Fig. 1, is a most simple plasmonic component, [24] and is easy to fabricate and integrate with ordinary passive plasmonic components in a plasmonic circuit. More importantly, the co-existence behavior is based on the interaction of modes at the band edge and in the intermode plasmonic band gap. [25-27] The band edge mode proceeds from the coupling of LSPs in a single resonator and the lattice diffraction. Different from the band edge mode in the ordinary grating geometry, the coupling of the band edge mode here is mediated by SP waves propagating on the silver film. Therefore, this grating-film structure can act as out-couplers from the coherent plasmon source to in-plane SPs due to the continuous feature of the silver film. And at the same time, the coherent plasmon also can be transformed to directional emitted photons since the grating periodicity can compensate the momentum mismatch between photons and SPs. Compared with usual localized SP modes, modes at the band edge and in the plasmonic band gap could significantly decrease the spasing threshold power. By carefully designing parameters, we demonstrate the dynamic amplification and co-existence of lasing mode (the band edge mode) and spasing mode (mode in the plasmonic band gap). Moreover, we investigate the mode competition and suppression of spaser and lasing spaser. We also show that the field enhancement of spaser mode and the vertical emission of lasing spaser mode.

For simplicity and to obtain insight into the nature of the mode competition, here we only consider one- dimensional case. Figure 1 shows the schematic of the adopted nanostructure with a simple metal grating-film configuration. It is composed of (from bottom to top) a quartz substrate, a silver film (200 nm thick), a one-dimensional silver grating (100 nm thick) and an active layer PMMA doping with Rhodamine B (400 nm thick). The grating grid is 100 nm and the period is 460 nm. Optical gain can be obtained from the active layer by external pumping (e.g., a frequency-double Nd: yttrium aluminum garnet laser with wavelength of 532nm, the pulse duration 8 nanosecond and the power of 100 millijoule per pulse; the focused intensity can be reached to $1.62\times 10^{10}$ W/cm$^2$). SPs supported at PMMA-grating interface are amplified by dyes in PMMA adjacent to the silver grating, which provides the excitation of SPs and coherent optical feedback to SPs, and decouples the localized SPs confined by the grating surface into the free space.

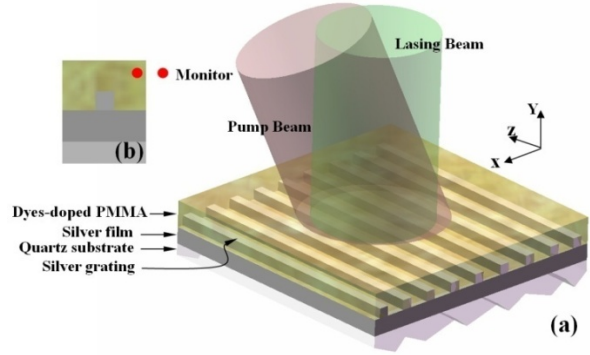

FIG. 1. (Color online) Schematic of the adopted structure.

Our work is based on a self-consistent Maxwell-Bloch (MB) approach, which deals with time-dependent wave equations by coupling the Maxwell equations with the rate equations of electron populations.[28-31] The advantages of this self-consistent approach are that one can follow the dynamical evolution process of the amplified electric field inside the system, and the amplification is nonlinear and saturated. The active medium, dye molecules (Rhodamine B), are described by a four-level atomic system in semiclassical theory.[28] The population density in each level is given by $N_0$, $N_1$, $N_2$ and $N_3$, respectively. The Maxwell equations for isotropic media are given by $\nabla\times\mathbf{E}=-\mu\mu_0\partial\mathbf{H}/\partial t$ and $\nabla\times\mathbf{H}=\varepsilon\varepsilon_0\partial\mathbf{E}/\partial t+\partial\mathbf{P}/\partial t$, where $\mathbf{P}$ is the electric polarization density that introduces gain and corresponds to the transitions between two atomic levels, $N_1$ and $N_2$. One can show that $\mathbf{P}$ obeys the following equation of motion[28]

$$\frac{\partial^2 \mathbf{P}(t)}{\partial t^2} + \Delta\omega \frac{\partial \mathbf{P}(t)}{\partial t} + \omega_a^2 \mathbf{P}(t) = \sigma_a \Delta N(t) E(t), \tag{1}$$

where $\omega_a$ is the center frequency of radiation and taken to be $2\pi \times 4.829 \times 10^{14}$ Hz (621 nm). $\Delta\omega = 2\pi \times 8 \times 10^{13}$ is the linewidth of the atomic transitions at $\omega_a$ and accounts for both dephasing processes and the nonradiative energy decay rate. $\sigma_a = (\gamma_r / \gamma_c) \cdot (e^2 / m)$ is the coupling strength of **P** to the external electric field, and its value is taken to be 9.77×10-5 C²/kg.

The atomic population densities obey the following rate equations[29-31]:

$$\frac{\partial N_3(r,t)}{\partial t} = \Gamma_{pump} N_0(r,t) - \frac{N_3(r,t)}{\tau_{32}}, \tag{2.1}$$

$$\frac{\partial N_2(r,t)}{\partial t} = \frac{N_3(r,t)}{\tau_{32}} + \frac{1}{\hbar\omega_a} \mathbf{E}(r,t) \cdot \frac{\partial \mathbf{P}(r,t)}{\partial t} - \frac{N_2(r,t)}{\tau_{12}}, \tag{2.2}$$

$$\frac{\partial N_1(r,t)}{\partial t} = \frac{N_2(r,t)}{\tau_{12}} - \frac{1}{\hbar\omega_a} \mathbf{E}(r,t) \cdot \frac{\partial \mathbf{P}(r,t)}{\partial t} - \frac{N_1(r,t)}{\tau_{10}}, \tag{2.3}$$

$$\frac{\partial N_0(r,t)}{\partial t} = \frac{N_1(r,t)}{\tau_{10}} - \Gamma_{pump} N_0(r,t), \tag{2.4}$$

where the pumping rate $\Gamma_{pump}$ is equivalent to a pump intensity.[32] Based on real laser dyes Rhodamine B, the lifetimes of the four energy levels, $\tau_{32}$, $\tau_{21}$, and $\tau_{10}$, are chosen to be $5\times10^{-14}$ s, $5\times10^{-12}$ s, and $5\times10^{-14}$ s, respectively. The total electron density $N_{total}(t=0) = N_0(t) + N_1(t) + N_2(t) + N_3(t)$ is taken to be $3.311 \times 10^{24} / m^3$. This approach is implemented by the finite-difference time-domain technique.[29-31]

First, we investigate the optical property of the system without gain. The structure is excited by a Gaussian pulse source (Fig. 2a) centered at a frequency of interest at a spatial point or having a spatial distribution in the form of a plane wave. We examine the temporal evolution of the electric field inside and just outside the system at the different points (Fig. 2b). And then the emission spectra and the surface plasmon resonance modes inside the system (Fig. 2c) are obtained by Fast Fourier Transformation (FFT). One can see that there exist three cold-cavity modes A (band edge mode), B (mode in the plasmonic band gap) and C (band edge mode).[26] The corresponding quality factors $\lambda/\Delta\lambda$ are 18.43, 248.54 and 29.42, respectively. The reflection is plotted in Fig. 2d, and here we want to emphasize that there are only two valleys which consist with modes A and C shown in Fig. 2c. Therefore, the cold-cavity modes A and C of the passive structure are bright modes of lasing spaser around 579.2 nm and 670 nm. Mode B is a dark mode of the spaser at 596.8 nm and does not couple to the far-zone optical field, so it is missing in Fig. 2d. It should be noted that the quality factor Q=248.54 of mode B is higher than Q=10~100 of the dark mode of the usual spaser.[2]

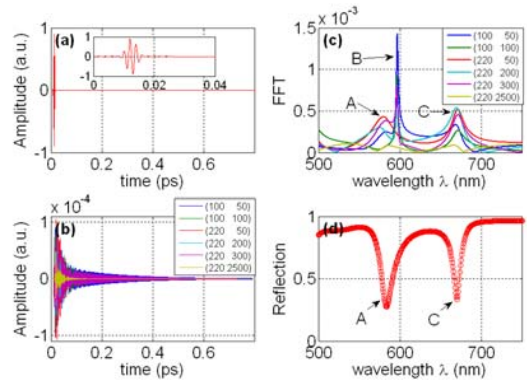

FIG. 2. (Color online) (a) The Gaussian pulse source inside the nanostructure. (b) The temporal evolution of fields at different points inside and just outside the nanostructure. (c) FFT of fields in Fig. 2b. (d) Reflection of the nanostructure.

Then, we investigate the lasing characteristic of modes in such system. The electric field at some point versus time for $\Gamma_{pump}=1\times10^{10}s^{-1}$ (Fig. 3a) shows that it is firstly amplified, then over-saturated and finally becomes steady after a long relaxation time. And the steady electric distribution after 1.5 picoseconds for $\Gamma_{pump}=1\times10^{10}s^{-1}$ is shown in Fig. 3d, showing that the spaser exists near the silver grid. From Fig. 3b, one can also see that it lases with one single mode with the wavelength of 596.8 nm for different pumping rates $\Gamma_{pump}=1\times10^{8}s^{-1}$, $1\times10^{9}s^{-1}$, $1\times10^{10}$s-¹. And when the gain is really big, e.g., for the pumping rates $\Gamma_{pump}=2\times10^{11}$s-¹ and $\Gamma_{pump}=1\times10^{12}$s-¹, it lases with two modes. The lasing frequencies are equal to that of the resonant modes B and C without gain (Fig. 2c). The fact that mode B lases with lower pumprate (Fig. 3b) indicates that it has a lower threshold than mode C. It agrees that mode B has the higher quality factor of 248.54 than mode C. And for $\Gamma_{pump}=1\times10^{12}$s-¹, the steady electric distributions (their phase difference is 0.25 cycle) after 1.5 picoseconds are shown in Fig. 3e and Fig. 3f. These distributions prove that there exist two different modes: one is similar with Fig. 3d and the other emits to the far field. To better understand the threshold behaviors of mode B and C, we extract the steady-state electric field intensity at mode B and C with increasing pump rates. The input-output curves for mode B, log-log plotted in Fig. 3c, shows a well-defined lasing threshold ($\Gamma_{pump}=2\times10^{7}$ s⁻¹) and linear dependence on the pump rate above and below threshold. And for mode C, a higher pumping rate ($\Gamma_{pump}=1.6\times10^{11}$ s⁻¹) presents and an electronic field amplified region exists between the two pumping rates.

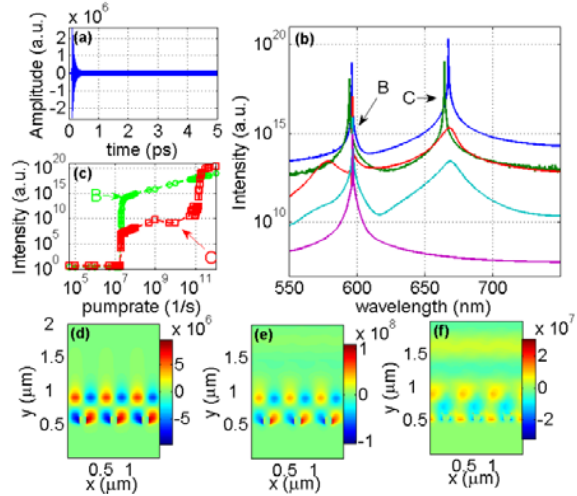

FIG. 3. (Color online) (a) The electronic field Ex versus time. (b) The intensity spectrum for the different pumping rates. From the below to the above: $\Gamma_{pump}=1\times10^{8}$s-¹, $1\times10^{9}$s-¹, $1\times10^{10}$s-¹, $2\times10^{11}$s-¹, $1\times10^{12}$s-¹. (c) The intensity at mode B and C versus the pumping rate. (d) The steady

electric distribution after 1.5 picoseconds for $\Gamma_{pump}=1\times10^{10}s^{-1}$. (e) The steady electric distributions after 1.5 picoseconds for $\Gamma_{pump}=1\times10^{12}s^{-1}$. (f) The steady electric distributions after 1.5 picoseconds for $\Gamma_{pump}=1\times10^{12}s^{-1}$. The phase difference between (e) and (f) is 0.25 cycle.

To obtain insight into the nature of the co-existence phenomena of modes B and C, we now calculate their steady-state electric field at their own resonant frequencies for the same pump rate. It is presented that both modes with wavelengths of 596.6 nm (Fig. 4b: mode B) and 667.7 nm (Fig. 4c: mode C) concentrate their energy at different places. Therefore, at different places, their strong electric fields force the electrons of the upper level $N_2$ to jump down to the $N_1$ level by stimulated emission. This leaves enough upper electrons for stimulated emissions of each other and then they can both exist in the same system.

The electronic field distribution shown in Fig. 4b also shows that mode B doesn't emit photons, so it is the dark mode of spaser. Moreover, it looks very strong and in fact, it is three times of the radiation field of mode C shown in Fig. 4c. This agrees with the above results that it has lower threshold and higher quality factor Q=248.54 of the corresponding cold-cavity mode. Therefore, it may be more suitable for the surface enhanced Raman scattering than the dark mode of spaser based on nanosphere, whose quality factor is Q=10~100.[2] In addition, the dark mode can propagate along the silver surface since it results from the coupling of the grating diffraction and LSP. It is obvious that this feature is useful in SP source.

From Fig. 4c, one can clearly observe that the lasing of mode C at the normal direction to the grating plane from the field distributions. It is shown that this mode is the bright mode of lasing spaser and that is to say, coherent light emission is possible in the nanoplasmonic gain-enhanced grating structure despite the presence of dark modes, which may deplete the gain before any bright mode can cross the threshold. Moreover, the bright mode emits vertically from large-area grating surfaces and therefore can be able to mold the spectral and angular distribution of the emission in solid-state lighting.

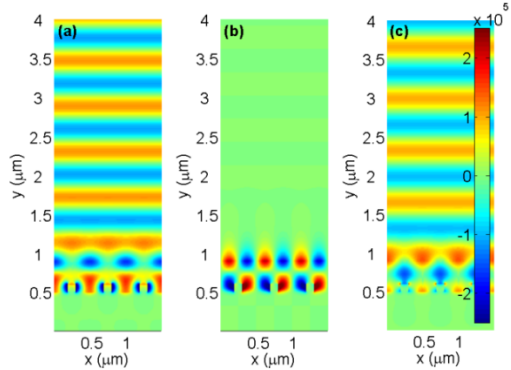

FIG. 4. (Color online) The distribution of the electronic field Ex at wavelength of (a) 582.3 nm, (b) 596.6 nm and (c) 667.7 nm.

In Fig. 3b, it is also shown that the number of lasing modes increases with increasing the pumping rate. We find that mode A is missing and the number of lasing modes is less than that of cold-cavity modes even for very large gain ($\Gamma_{pump}=1\times10^{12}$s$^{-1}$, far above the threshold). These multi-lasing peaks and the saturated-mode-number phenomena can be attributed to the interplay between spatial-hole-burning and amplification.

Finally, to further understand the saturated-mode-number phenomena, we examine the steady-state electric field of mode A and B for the same pump rate. Both mode A at wavelength of 582.3 nm (Fig. 4a) and mode B at wavelength of 596.6 nm (Fig. 4b) focus most of their energy near the grating grid. That is to say, mode B partly overlaps A in the whole system, and its strong electric field can force almost all the electrons of the upper level $N_2$ to jump down to the $N_1$ level quickly by stimulated emission. This leaves very few upper electrons for stimulated emission of mode A. In other words, mode A is suppressed by the first lasing mode even though its threshold value is only a little bit bigger than the first one. This mode suppression phenomenon due to spatial hole burning also exists in common homogeneously broadening lasers.[28]

In summary, based on a simple and reliable grating-film geometry, we simultaneously realize spaser and vertically emitted plasmon laser. We demonstrate three cold-cavity modes in our system: one is dark mode (does not radiate) and others are bright modes. Comparing to the dark mode in a spaser based LSP of one nanoparticle (Q=10-100),[1] the dark mode in our system has higher quality factor (Q=248.54). Moreover, and also the most important, coherent light emits vertically from the gain-enhanced grating-film structure despite the presence of dark modes, which may deplete the gain before any bright mode can cross the threshold. Both of them simultaneously lase and compete in the frequency regime we focus on. In addition, the lasing mode number increased with the pumped power, while the number of lasing modes is less than that of cold-cavity modes due to gain competition of different modes and spatial hole burning. The demonstrated phenomena suggest that large-area laser and spaser can be realized in a single platform. Our results imply a broad possibility of applications such as low threshold SP sources, surface enhanced Raman scattering, solid-state lighting emission, and so on.

This work was supported by the National Natural Science Foundation of China (grants 61107032, 61275170 and 51003113). The authors would like to thank engineer Wenda Han of Dongjun Information Technology Co., Ltd. for his efforts on the customized version of eastFDTD software.

**References**

[1] D. J. Bergman and M. I. Stockman, Phys. Rev. Lett. **90,** 027402 (2003).

[2] Stockman, M. I., Nat. Photonics **2**, 327–329 (2008).

[3] M. A. Noginov, G. Zhu, A. M. Belgrave, R. Bakker, V. M. Shalaev, E. E. Narimanov, S. Stout, E. Herz, T. Suteewong, and U. Wiesner, Nature **460**, 1110 (2009).

[4] Martin T. Hill, Milan Marell, Eunice S. P. Leong, Barry Smalbrugge, Youcai Zhu, Minghua Sun, Peter J. van Veldhoven, Erik Jan Geluk, Fouad Karouta, Yok-Siang Oei, Richard Nötzel, Cun-Zheng Ning, and Meint K. Smit, Opt. Express **17,** 11107-11112 (2009).

[5] R.F. Oulton, V.J. Sorger, T. Zentgraf, R.-M. Ma, C. Gladden, L. Dai, G. Bartal, and X. Zhang, Nature **461,** 629 (2009).

[6] R. M. Ma, R.F. Oulton, V.J. Sorger, G. Bartal, and X. Zhang, Nat. Mater. **10(2),** 110–113 (2011).

[7] Chen-Ying Wu, Cheng-Tai Kuo, Chun-Yuan Wang, Chieh-Lun He, Meng-Hsien Lin, Hyeyoung Ahn, and Shangjr Gwo, Nano Lett., **11,** 4256–4260 (2011).

[8] Yu-Jung Lu, Jisun Kim, Hung-Ying Chen, Chihhui Wu, Nima Dabidian, Charlotte E. Sanders, Chun-Yuan Wang, Ming-Yen Lu, Bo-Hong Li, Xianggang Qiu, Wen-Hao Chang, Lih-Juann Chen, Gennady Shvets, Chih-Kang Shih, Shangjr Gwo, Science **337,** 450 (2012).

[9] K. Ding, Z. C. Liu, L. J. Yin, M. T. Hill, M. J. H. Marell, P. J. van Veldhoven, R. N¨oetzel, and C. Z. Ning, Phys. Rev. B **85,** 041301 (2012).

[10] Kang Ding, Leijun Yin, Martin T. Hill, Zhicheng Liu, Peter J. van Veldhoven and C. Z. Ning, Appl. Phys. Lett. **102,** 041110 (2013).

[11] E. S. Andrianov, A. A. Pukhov, A. V. Dorofeenko, A. P. Vinogradov, and A. A. Lisyansky, Opt. Express **19,** 24849 (2011).

[12]E. S. Andrianov, A. A. Pukhov, A. V. Dorofeenko, A. P. Vinogradov, and A. A. Lisyansky, Phys. Rev. B **85,** 165419 (2012).

[13]V. M. Parfenyev and S. S. Vergeles, Phys. Rev. A **86,** 043824 (2012).

[14]E. S. Andrianov, A. A. Pukhov, A. V. Dorofeenko, and A. P. Vinogradov and A. A. Lisyansky, Phys. Rev. B **85,** 035405 (2012).

[15]Pavel Ginzburg and Anatoly V. Zayats, Opt. Express **21,** 2147 (2013).

[16]Dabing Li and Mark I. Stockman, Phys. Rev. Lett. **110,** 106803 (2013).

[17]Xiangeng Meng, Urcan Guler, Alexander V. Kildishev, Koji Fujita, Katsuhisa Tanaka & Vladimir M. Shalaev, Scientific Reports, **3,** 1241 (2013).

[18]M. I. Stockman, Phys. Rev. Lett. **106,** 156802 (2011).

[19]Y. L. Zhang, W. Jin, X. Z. Dong, Z. S. Zhao and X. M. Duan, Opt. Express 20(10), 10776 (2012).

[20]O. Hess, J. B. Pendry, S. A. Maier, R. F. Oulton, J. M. Hamm and K. L. Tsakmakidis, Nat. Mater. **11,** 573 (2012).

[21]N. I. Zheludev, S. L. Prosvirnin, N. Papasimakis, and V. A. Fedotov, Nat. Photonics **2(6),** 351 (2008).

[22]V. A. Fedotov, N. Papasimakis, E. Plum, A. Bitzer, M. Walther, P. Kuo, D. P. Tsai, and N. I. Zheludev, Phys. Rev. Lett. **104,** 223901 (2010).

[23]Yao-Wei Huang, Wei Ting Chen, Pin Chieh Wu, Vassili A. Fedotov2, Nikolay I. Zheludev & Din Ping Tsai, Scientific Reports, **3,** 1237 (2013).

[24]V.M.Shalaev and S. Kawata, Nanophotonics with Surface Plasmons (Elsevier, 2007).

[25]D.Zhao, C.Zhou, Y.Zhang, L.Shi and X.Jiang, Appl. Phys. B. **91,** 475 (2008).

[26]Takayuki Okamoto, Janne Simonen, and Satoshi Kawata, Phys. Rev. B. **77,** 115425 (2008).

[27]X. F. Li,a_ S. F. Yu,b_ and Ashwani Kumar, Appl. Phys. Lett. **95,** 141114 (2009).

[28] A. E. Siegman, Lasers (University Science, Sausalito, CA, 1986). See Chaps. 3, 8, and 9.

[29] A. Fang, T. Koschny, M. Wegener, and C. M. Soukoulis, Phys. Rev. B **79,** 241104 (2009).

[30] S. Wuestner, A. Pusch, K. L. Tsakmakidis, J. M. Hamm and O. Hess, Phil. Trans. R. Soc. A **369**, 3525 (2011).

[31] http://www.eastfdtd.com/index.asp#.

[32] The pumping rate $\Gamma_{pump}$ denotes there are $\Gamma_{pump} \cdot N_0(t)$ electrons excited from $N_0$ level to $N_3$ level per second, and the corresponding excitation intensity in W/m$^3$ is equal to $\Gamma_{pump} \cdot N_0 \cdot h\omega/(2\pi)$. Thus the excitation intensity in W/cm$^2$ is $\Gamma_{pump} \cdot N_0 \cdot h\omega/(2\pi) \cdot$ (volume/surface area). Taking into account the pump wavelength 532 nm and the thickness 400 nanometer of the gain layer, the excitation intensity in W/cm$^2$ is $P = \Gamma_{pump} \cdot N_0 \cdot hc/(532 \times 10^{-9}) \cdot 10^{-2} \cdot 10^{-2} \cdot 400 \cdot 10^{-9}$. For the steady state, the electrons at $N_0$ level is $N_0 = N_{total} \cdot (1 - \tau_{21} \cdot P_{pump}/(1 + \tau_{21} \cdot P_{pump}))$. Here $N_{total}$ is $3.311 \times 10^{24}$/m$^3$ and $\tau_{21} = 5 \times 10^{-12}$s, for a typical pumping rate $\Gamma_{pump} = 1 \times 10^{10}$s$^{-1}$, the excitation intensity is $4.7097 \times 10^5$ W/cm$^2$.